\documentclass[10pt,journal]{IEEEtran}

\makeatletter
\def\ps@headings{%
\def\@oddhead{\mbox{}\scriptsize\rightmark \hfil \thepage}%
\def\@evenhead{\scriptsize\thepage \hfil\leftmark\mbox{}}%
\def\@oddfoot{}%
\def\@evenfoot{}}
\makeatother
\usepackage{latexsym}
\usepackage{amssymb}
\usepackage{epsfig}
\usepackage{amsthm}
\usepackage{subfigure}
\usepackage{graphics,color,psfrag}
\usepackage{graphicx}
\usepackage{amsmath}
\usepackage{algorithm}
\usepackage{algorithmic}
\usepackage{cite}
\usepackage{comment}
\usepackage{url}
\usepackage{xspace}

\usepackage{xspace}
\usepackage{bbm}
\input{mathlig}

\newcommand{\muspace}{\mspace{1mu}}

\DeclareRobustCommand{\scond}{\mathchoice{\muspace\vert\muspace}{\vert}{\vert}{\vert}}
\mathlig{|}{\scond}

\DeclareRobustCommand{\discint}{\mathchoice{\mspace{-1.5mu}:\mspace{-1.5mu}}{\mspace{-1.5mu}:\mspace{-1.5mu}}{:}{:}}
\mathlig{::}{\discint}

\newcommand{\Bc}{\mathcal{B}}
\newcommand{\Cc}{\mathcal{C}}
\newcommand{\cc}{\text{\footnotesize $\mathcal{C}$} }

\newcommand{\Ec}{\mathcal{E}}

\newcommand{\Sc}{\mathcal{S}}

\newcommand{\Uc}{\mspace{1.5mu}\mathcal{U}}

\newcommand{\pen}{{P_e^{(n)}}}

\newcommand{\aep}{{\mathcal{T}_{\epsilon}^{(n)}}}
\newcommand{\aepvar}{{\mathcal{T}_{\epsilon'}^{(n)}}}

\newcommand{\Mh}{{\hat{M}}}

\newcommand{\Yh}{{\hat{Y}}}

\newcommand{\lh}{{\hat{l}}}
\newcommand{\mh}{{\hat{m}}}

\newcommand{\Rt}{{\tilde{R}}}

\newcommand{\Ut}{{\tilde{U}}}

\newcommand{\Xt}{{\tilde{X}}}
\newcommand{\Yt}{{\tilde{Y}}}

\newcommand{\ut}{{\tilde{u}}}

\newcommand{\xt}{{\tilde{x}}}

\def\d{\delta}
\def\e{\epsilon}

\let\P\relax
\DeclareMathOperator\P{\textsf{P}}

\def\textiid{i.i.d.\@\xspace}
\newcommand\iid{\ifmmode\text{ i.i.d. } \else \textiid \fi}

\def\clap#1{\hbox to 0pt{\hss#1\hss}}
\def\mathclap{\mathpalette\mathclapinternal}
\def\mathclapinternal#1#2{%
  \clap{$\mathsurround=0pt#1{#2}$}}

\let\oldstackrel\stackrel
\renewcommand{\stackrel}[2]{\oldstackrel{\mathclap{#1}}{#2}}

\def\x{{\mathbf x}}

\newtheorem{theorem}{Theorem}
\newtheorem{lemma}{Lemma}

\theoremstyle{definition}

\newtheorem{remarks} {Remark}
\newtheorem{example}{Example}
\newtheorem{claim}{Claim}

\renewcommand{\qed}{\nobreak \ifvmode \relax \else
      \ifdim\lastskip<1.5em \hskip-\lastskip
      \hskip1.5em plus0em minus0.5em \fi \nobreak
      \vrule height0.2em width0.5em depth0.4em\fi}
\IEEEoverridecommandlockouts
\begin{document}

\title{On the Capacity of the Noncausal Relay Channel}
\author{Lele Wang~\IEEEmembership{Member,~IEEE}, and Mohammad Naghshvar~\IEEEmembership{Member,~IEEE}
\thanks{This paper was presented in part at IEEE International Symposium on Information Theory 2011.}
\thanks{L. Wang is jointly with the Department of Electrical Engineering, Stanford University, Stanford, CA, USA and the Department of Electrical Engineering - Systems, Tel Aviv University, Tel Aviv, Israel (email: wanglele@stanford.edu).}
\thanks{M. Naghshvar was with the Department of Electrical and Computer Engineering, University of California San 
Diego, La Jolla, CA 92093 USA. He is now with Qualcomm Technologies Inc., San Diego, CA 
92121 USA (e-mail: mnaghshv@qti.qualcomm.com).}}
\maketitle

\begin{abstract}
This paper studies the noncausal relay channel, also known as the relay channel
with unlimited lookahead, introduced by El Gamal, Hassanpour, and Mammen. Unlike
the standard relay channel model, where the relay encodes its signal based on
the previous received output symbols, the relay in the noncausal relay channel
encodes its signal as a function of the entire received sequence. In the
existing coding schemes, the relay uses this noncausal information solely to
recover the transmitted message or part of it and then cooperates with the
sender to communicate this message to the receiver. However, it is shown in this
paper that by applying the Gelfand--Pinsker coding scheme, the relay can take
further advantage of the noncausally available information and achieve rates
strictly higher than those of the existing coding schemes. This paper also
provides a new upper bound on the capacity of the noncausal relay channel that
strictly improves upon the existing cutset bound. These new lower and upper
bounds on the 
capacity coincide for the class of degraded noncausal relay channels and
establish the capacity for this class.
\end{abstract}


\section{Introduction}

The relay channel was first introduced by van der Meulen \cite{Meulen}. In their
classic paper \cite{Cover79}, Cover and El Gamal established the cutset upper
bound and the decode--forward, partial decode--forward, compress--forward, and
combined lower bounds
for the relay channel. Furthermore, they established the capacity for the
classes of degraded and reversely degraded relay channels, and relay channels
with feedback.

The relay channel with lookahead was introduced by El~Gamal, Hassanpour, and
Mammen~\cite{Hassanpour07}, who mainly studied the following two classes:
\begin{itemize}
\item Causal relay channel (also known as \emph{relay-without-delay}) in which
the relay has access only to the past and present received sequence. This model
is usually considered when the delay from the sender-receiver link is
sufficiently longer than the delay from the sender-relay link so that the relay
can depend on the ``present'' in addition to the past received sequence. A lower
bound for the capacity of this channel was established by combining partial
decode--forward and instantaneous relaying coding schemes. The cutset upper
bound for the causal relay channel was also established.

\item Noncausal relay channel (also known as
\emph{relay-with-unlimited-lookahead}) in which the relay knows its entire
received sequence in advance and hence the relaying functions can depend on the
whole received block. This model provides a limit on the extent to which
relaying can help communication. Lower bounds on the capacity were established
by extending (partial) decode--forward coding scheme to the noncausal case. The
cutset upper bound for the noncausal relay channel was also established.

\end{itemize}

The focus of this paper is on the noncausal relay channel. The existing lower
bounds on the capacity of this channel are derived using the (partial)
decode--forward coding scheme. In particular, the relay recovers the transmitted
message from the received sequence (available noncausally at the relay) and then
cooperates with the sender to coherently transmit this message to the receiver.
Therefore, the noncausally available information is used solely to recover the
transmitted message at the relay. However, it is shown in this paper that the
relay can take further advantage of the received sequence by considering it as
noncausal side information to help the relay's communication to the receiver.
Based on this observation, we establish in this paper several improved lower
bounds on the capacity of the noncausal relay channel by combining the
Gelfand--Pinsker coding scheme~\cite{GelfandPinsker} with (partial)
decode--forward and compress--forward at the relay.
Moreover, we establish a new upper bound on the capacity that improves upon the
cutset bound~\cite[Theorem 16.6]{Kim}. The new upper bound is shown to be tight
for the class of degraded noncausal relay channels and is achieved by the
Gelfand--Pinsker decode--forward coding scheme.

The rest of the paper is organized as follows. In
Section~\ref{ProblemFormulation}, we formulate the problem and provide a brief
overview of the existing literature. In Section~\ref{LowerBounds}, we establish
three improved lower bounds, the Gelfand--Pinsker decode--forward (GP-DF) lower
bound, the Gelfand--Pinsker compress--forward lower bound, and the
Gelfand--Pinsker partial decode--forward compress--forward lower bound. We show
through Example~\ref{ex:BEC} that the GP-DF lower bound can be strictly tighter
than the existing lower bound. In Section~\ref{UpperBound}, we establish a new
upper bound on the capacity, which is shown through Example~\ref{ex:CSvsNUB} to
strictly improve upon the cutset bound. This improved upper bound together with
the GP-DF lower bound establishes the capacity for the class of degraded
noncausal relay channels.

\begin{figure*}[t]
\centering
\small
\psfrag{X1}{$X_{1i}$}
\psfrag{p}[cc]{{$p(y_2|x_1)p(y_3|x_1,x_2,y_2)$}}
\psfrag{Y2}{$Y_{2i}$}
\psfrag{X2}{$X_{2i}(Y_2^{i+d})$}
\psfrag{Y3}{$Y_{3i}$}
\psfrag{E}[cc]{Encoder}
\psfrag{D}[cc]{Decoder}
\psfrag{RE}[cc]{\;Relay Encoder}
\psfrag{M}{$M$}
\psfrag{Mh}{$\hat{M}$}
\includegraphics[width=.7\textwidth]{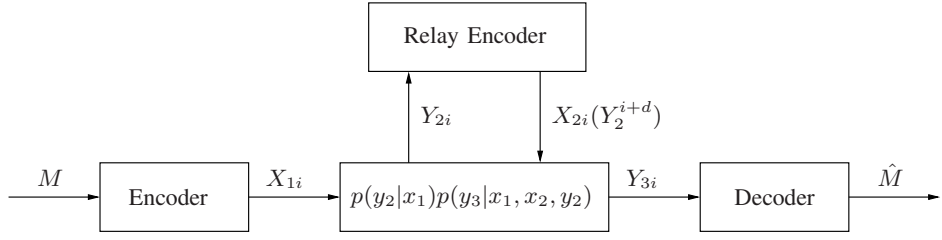}
\caption{Relay channel with lookahead $d \in \mathbb{Z}$}
\label{LARelay}
\end{figure*}

Throughout the paper, we follow the notation in~\cite{Kim}. In particular, a
random variable is denoted by an upper case letter (e.g., $X,Y,Z$) and its
realization is denoted by a lower case letter (e.g., $x,y,z$). By convention,
$X=\emptyset$ means that $X$ is a degenerate random variable (unspecified
constant) regardless of its support. Let  $X_k^n=(X_{k1},X_{k2},\ldots,X_{kn})$.
We say that $X \to Y \to Z$ form a Markov chain if $p(x,y,z)=p(x)p(y|x)p(z|y)$.
For $a \ge 0$, $[1:2^a]=\{1,2,\ldots,2^{\left\lceil a \right\rceil}\}$, where
$\left\lceil a \right\rceil$ is the smallest integer greater than or equal to
$a$. For any set $\Sc$, $\left| \Sc \right|$ denotes its cardinality. The
probability of an event $\mathcal{A}$ is denoted by $\P(\mathcal{A})$.


\section{Problem Formulation and Known Results}
\label{ProblemFormulation}

\subsection{Noncausal Relay Channel}

Consider the $3$-node point-to-point communication system with a relay depicted
in Figure \ref{LARelay}. The sender (node $1$) wishes to communicate a message
$M$ to the receiver (node $3$) with the help of the relay (node $2$). The {\it
discrete memoryless (DM) relay channel with lookahead} is described as
\begin{align}
\label{DMLA}
(\mathcal{X}_1 \times \mathcal{X}_2,p(y_2|x_1)p(y_3|x_1,x_2,y_2),\mathcal{Y}_2
\times \mathcal{Y}_3,d)
\end{align}
where the parameter $d \in \mathbb{Z}$ specifies the amount of lookahead. The
channel is memoryless in the sense that $p(y_{2i}|x_1^i,y_2^{i-1},m) =
p_{Y_2|X_1}(y_{2i}|x_{1i})$ and $p(y_{3i}|x_1^i,x_2^i,y_2^i,y_3^{i-1},m) =
p_{Y_3|X_1,X_2,Y_2}(y_{3i}|x_{1i},x_{2i},y_{2i})$.

A $(2^{nR},n)$ code for the relay channel with lookahead consists of
\begin{itemize}
\item a message set $[1::2^{nR}]$,
\item an encoder that assigns a codeword $x_1^n(m)$ to each message $m \in
[1::2^{nR}]$,
\item a relay encoder that assigns a symbol $x_{2i}(y_2^{i+d})$ to each sequence
$y_2^{i+d}$ for $i \in [1::n]$, where the symbols that have nonpositive time
indices or time indices greater than $n$ are arbitrary, and
\item a decoder that assigns an estimate $\mh(y_3^n)$ or an error message $e$ to
each received sequence $y_3^n$.
\end{itemize}

We assume that the message $M$ is uniformly distributed over $[1:2^{nR}]$. The
average probability of error is defined as $\pen = \P\{\Mh \ne M\}$. A rate $R$
is said to be {\it achievable} for the DM relay channel with lookahead if there
exists a sequence of $(2^{nR},n)$ codes such that $\lim_{n\rightarrow \infty}
\pen = 0$. The capacity $C_{d}$ of the DM relay channel with lookahead is the
supremum of all achievable rates.

The standard DM relay channel\footnote[1]{Note that here we define the relay
channel with lookahead as $p(y_2|x_1)p(y_3|x_1,x_2,y_2)$,
since the conditional pmf $p(y_2,y_3|x_1,x_2)$ depends on the code due to the
instantaneous or lookahead dependency of $X_2$ on $Y_2$.}
corresponds to lookahead parameter $d = -1$, or equivalently, a delay of $1$.
Causal relay channel corresponds to lookahead parameter $d = 0$, i.e., the
relaying function at time $i$ can depend only on the past and present relay
received sequence $y_2^i$ (instead of $y_2^{i-1}$ as in the standard relay
channel). The noncausal relay channel which we focus on in this paper is the
case where $d = \infty$, i.e., the relaying functions can depend on the entire
received sequence $y_2^n$. The purpose of studying this extreme case is to
quantify the limit on the potential gain from relaying.

\subsection{Prior Work}
The noncausal relay channel was initially studied by El~Gamal, Hassanpour, and
Mammen~\cite{Hassanpour07}, who established the following lower bounds on the
capacity $C_{\infty}$. The cutset upper bound on the capacity $C_\infty$ is due
to~\cite[Theorem 16.6]{Kim}.
\begin{itemize}
\item Decode--forward (DF) lower bound:
	\begin{align}	
	C_{\infty} &\ge R_{\textrm{DF}}\nonumber\\
	&= \max_{p(x_1,x_2)} \min \left\{
I(X_1;Y_2) ,I(X_1,X_2;Y_3) \right\}. \label{DecLwBnd}
	\end{align}
\item Partial decode--forward (PDF) lower bound:
	\begin{align}
	C_{\infty} &\geq R_{\textrm{PDF}} \nonumber\\
	&=  \max \limits_{p(v,x_1,x_2)} \min \{
	 I(V;Y_2)+I(X_1;Y_3|X_2,V),\nonumber\\
	 &\hspace{7.2em}I(X_1,X_2;Y_3)\}.\label{PDecLwBnd}
	\end{align}
\item Cutset bound\footnote[2]{There is a small typo
in \cite[Theorem 1]{Hassanpour07} where the maximum is over $p(x_1,x_2)$
instead of $p(x_1)p(x_2|x_1,y_2)$.} for the noncausal relay channel:
\begin{align}
C_{\infty} & \leq R_{\textrm{CS}}\nonumber\\
&= \max
\limits_{{p(x_1)p(u|x_1,y_2)x_2(u,y_2)}} \min \{I(U,X_1;Y_3)\nonumber\\
&\hspace{5em}I(X_1;Y_2)+ I(X_1;Y_3|X_2,Y_2),\}.\label{CutUpBnd}
\end{align}
\end{itemize}
%


\section{Lower Bounds}
\label{LowerBounds}

In this section, we establish three lower bounds by considering the received
$y_2^n$ sequence at the relay as noncausal side information to help
communication. In Subsection~\ref{sec:DF}, we first establish the
Gelfand--Pinsker decode--forward (GP-DF) lower bound by incorporating
Gelfand--Pinsker coding with the decode--forward coding scheme. Then we show
that the GP-DF lower bound can be strictly tighter than the decode--forward
lower bound and achieve the capacity in Example~\ref{ex:BEC}. In
Subsection~\ref{sec:CF}, we establish the Gelfand--Pinsker compress--forward
(GP-CF) lower bound in two different ways, one via the Wyner--Ziv binning with
Gelfand--Pinsker coding and another via the recently developed hybrid coding
techniques~\cite{Lim2010}~\cite{Lim2011}~\cite{Kim2011}. In
Subsection~\ref{sec:DFCF}, we further combine the hybrid coding techniques with
the partial decode--forward coding scheme.

\subsection{Gelfand--Pinsker Decode--Forward Lower Bound}
\label{sec:DF}
We first incorporate Gelfand--Pinsker coding with the decode--forward coding
scheme.

\begin{theorem}[Gelfand--Pinsker decode--forward (GP-DF) lower bound]
The capacity of the noncausal relay channel is lower bounded as
\begin{align}
C_{\infty} &\geq R_{{\textrm{\em GP-DF}}} \nonumber\\
&= \max \min\{I(X_1;Y_2),\;
I(X_1,U;Y_3)-I(U;Y_2|X_1)\}, \label{DFGP}
\end{align}
where the maximum is over all pmfs $p(x_1)p(u|x_1,y_2)$ and functions
$x_2(u,x_1,y_2)$.
\label{thm:DFGP}
\end{theorem}

\begin{remarks}
Taking $U$ conditionally independent of $Y_2$ given $X_1$ and setting $X_2 = U$
reduces the GP-DF lower bound to the DF lower bound in (\ref{DecLwBnd}).
\end{remarks}

\begin{IEEEproof} The GP-DF coding scheme uses {\it multicoding} and joint
typicality encoding and decoding. For each message $m$, we generate a $x_1^n(m)$
sequence and a {\it subcodebook} $\mathcal{C}(m)$ of $2^{n\tilde{R}}$ $u^n(l|m)$
sequences. To send message $m$, the
sender transmits $x_1^n(m)$. Upon receiving $y_2^n$ noncausally, the relay first
finds a message estimate $\tilde{m}$. It then finds a
$u^n(l|\tilde{m}) \in \mathcal{C}(\tilde{m})$ that is jointly typical with
$(x_1^n(\tilde{m}),y_2^n)$
and transmits $x_2^n(u^n(l|\tilde{m}),x_1^n(\tilde{m}),y_2^n)$.
The receiver declares $\hat{m}$ to be the message estimate if
$(x_1^n(\hat{m}),u^n(l|\hat{m}),y_3^n)$ are jointly typical for some
$u^n(l|\hat{m})\in \mathcal{C}(\hat{m})$. We now provide the details of the
proof. 

\smallskip
{\it Codebook generation:} Fix $p(x_1)p(u|x_1,y_2)$ and $x_2(u,x_1,y_2)$ that
attain the lower bound. Randomly and independently generate $2^{nR}$ sequences
$x_1^n(m)$, each according to $\prod_{i=1}^n p_{X_1}(x_{1i}), m \in [1:2^{nR}]$.
For each message $m \in [1:2^{nR}]$, randomly and conditionally independently
generate $2^{n\tilde{R}}$ sequences $u^n(l|m)$, each according to $\prod_{i=1}^n
p_{U|X_1}(u_i|x_{1i}(m))$, which form the subcodebook $\mathcal{C}(m)$. This
defines the codebook
$\mathcal{C}=\{(x_1^n(m),u^n(l|m),x_2^n(u^n(l|m),x_1^n(m),y_2^n)) :
m\in[1:2^{nR}], l\in[1:2^{n\tilde{R}}]\}$. The codebook is revealed to all
parties.

\smallskip
{\it Encoding:} To send message $m$, the encoder transmits $x_1^n(m)$.

\smallskip
{\it Relay encoding:} Upon receiving $y_2^n$ noncausally, the relay first finds
the unique message $\tilde{m}$ such that $(x_1^n(\tilde{m}),y_2^n) \in
\mathcal{T}_{\epsilon'}^{(n)}$. Then, it finds a sequence $u^n(l|\tilde{m}) \in
\mathcal{C}(\tilde{m})$ such that $(u^n(l|\tilde{m}),x_1^n(\tilde{m}),y_2^n) \in
\mathcal{T}_{\epsilon'}^{(n)}$. If there is more than one such index, it selects one 
of them uniformly at random. If there is no such index, it selects an index from 
$[1::2^{n\Rt}]$ uniformly at random. The relay transmits $x_{2i} =
x_{2}(u_i(l|\tilde{m}),x_{1i}(\tilde{m}),y_{2i})$ at time
$i \in [1:n]$.

\smallskip
{\it Decoding:} Let $\epsilon > \epsilon'$. Upon receiving $y_3^n$, the decoder
declares that $\hat{m} \in [1:2^{nR}]$ is sent if it is the unique message such
that $(x_1^n(\hat{m}),u^n(l|\hat{m}),y_3^n) \in \mathcal{T}_{\epsilon}^{(n)}$
for some $u^n(l|\hat{m})\in \mathcal{C}(\hat{m})$; otherwise, it declares an
error. 

\smallskip
{\it Analysis of the probability of error:} We analyse the probability of error
averaged over codes.
Assume without loss of generality that $M = 1$. Let $\tilde{M}$ be the relay's
message estimate and let $L$ denote the index of the chosen
$U^n$ codeword for $\tilde{M}$ and $Y_2^n$. The decoder makes an error only if
one of the following events occur:
{\allowdisplaybreaks
\begin{align*}
\tilde{\mathcal{E}} &= \{\tilde{M} \neq 1\},\\
\tilde{\mathcal{E}}_1 &= \{(X_1^n(1),Y_2^n) \notin
\mathcal{T}_{\epsilon'}^{(n)}\},\\
\tilde{\mathcal{E}}_2 &= \{(X_1^n(m),Y_2^n) \in \mathcal{T}_{\epsilon'}^{(n)}
\;\textrm{for some}\; m \neq 1\},\\
\tilde{\mathcal{E}}_3 &= \{(U^n(l|\tilde{M}), X_1^n(\tilde{M}),Y_2^n) \notin
\mathcal{T}_{\epsilon'}^{(n)}\\
 &\hspace{12em}\textrm{for all}\; U^n(l|\tilde{M})\in \mathcal{C}(\tilde{M})\},\\
\mathcal{E}_1 &= \{(X_1^n(1),U^n(L|1),Y_3^n) \notin
\mathcal{T}_{\epsilon}^{(n)}\},\\
\mathcal{E}_2 &= \{(X_1^n(m),U^n(l|m),Y_3^n) \in \mathcal{T}_{\epsilon}^{(n)}\\
 &\hspace{8.3em}\textrm{for some}\; m\neq 1, U^n{(l|m)\in \mathcal{C}(m)}\}.
\end{align*}
}%
Thus, the probability of error is upper bounded as
\begin{align*}
\P(\mathcal{E}) &= \P\{\hat{M}\neq 1\}\\
&\leq \P(\tilde{\mathcal{E}} \cup \tilde{\mathcal{E}}_3 \cup \mathcal{E}_1 \cup
\mathcal{E}_2)\\
&\leq \P(\tilde{\mathcal{E}}) + \P(\tilde{\mathcal{E}}_3 \cap
\tilde{\mathcal{E}}^c) + \P(\mathcal{E}_1 \cap \tilde{\mathcal{E}}^c \cap
\tilde{\mathcal{E}}_3^c) + \P(\mathcal{E}_2)\\
&\leq \P(\tilde{\mathcal{E}}_1) + \P(\tilde{\mathcal{E}}_2)+
\P(\tilde{\mathcal{E}}_3 \cap \tilde{\mathcal{E}}^c) \\
&\hspace{1em}+\P(\mathcal{E}_1 \cap
\tilde{\mathcal{E}}^c \cap \tilde{\mathcal{E}}_3^c)
 + \P(\mathcal{E}_2).
\end{align*}
By the law of large numbers (LLN), the first term tends to zero as $n
\rightarrow \infty$. By the packing lemma \cite{Kim}, the second term tends to
zero as $n \rightarrow \infty$ if $R<I(X_1;Y_2)-\delta(\epsilon')$. Therefore,
$\P(\tilde{\mathcal{E}})$ tends to zero as $n \rightarrow \infty$ if
$R<I(X_1;Y_2)-\delta(\epsilon')$. Given $\tilde{\mathcal{E}}^c$, i.e.
$\{\tilde{M}=1\}$, by the covering lemma \cite{Kim}, the third term tends to
zero as $n \rightarrow \infty$ if $\tilde{R}>I(U;Y_2|X_1)+\delta(\epsilon')$. By
the conditional typicality lemma, the fourth term tends to zero as $n
\rightarrow \infty$. Finally, note that once $m$ is wrong, $U^n(l|m)$ is also
wrong. By the packing lemma, the last term tends to zero as $n \rightarrow
\infty$ if $R+\tilde{R} < I(X_1,U;Y_3)-\delta(\epsilon)$. Combining the bounds
and eliminating $\tilde{R}$, we have shown that $\P\{\hat{M}\neq 1\}$ tends to
zero as $n \rightarrow \infty$ if $R<I(X_1;Y_2)-\delta(\epsilon')$ and
$R<I(X_1,U;Y_3)-I(U;Y_2|X_1)-\delta'(\epsilon)
$ where $\delta'(\epsilon)=\delta(\epsilon)+\delta(\epsilon')$.
This completes the proof.
\end{IEEEproof}

\begin{remarks}
Unlike the coding schemes for the regular relay channel, we do not need block
Markov coding for the noncausal relay channel for the following two reasons.
First, from the channel statistics $p(y_2|x_1)$, $Y_2$ does not depend on $X_2$
and hence there is no need to make $x_1^n$ correlated with the previous block
$x_2^n$. Second, $y_2^n$ is available noncausally at the relay and hence the
signals from the sender and the relay arrive at the receiver in the same block.
\end{remarks}


The GP-DF lower bound can be strictly tighter than the DF lower bound as shown
in the following example.
\begin{example}
\label{ex:BEC}
Consider a degraded noncausal relay channel $p(y_2|x_1)p(y_3|x_1,x_2,y_2) =
p(y_2|x_1)p(y_3|x_2,y_2)$ depicted in Figure \ref{DFGPvsDF}.
The channel from the sender to the relay is a $\mathrm{BEC}(1/2)$ channel, while
the channel from the relay to the receiver is clean if $Y_2 \in \{0,1\}$ and
stuck at $1$ if $Y_2$ is an erasure.

\begin{figure}[htbp]
\centering
\footnotesize
\psfrag{X1}{$X_1$}
\psfrag{Y2}{$Y_2$}
\psfrag{X2}{$X_2$}
\psfrag{Y3}{$Y_3$}
\psfrag{S0}[cc]{$Y_2 = 0,1$}
\psfrag{S1}[cc]{$Y_2 = \mathrm{e}$}
\psfrag{p}{1/2}
\psfrag{0}{$0$}
\psfrag{1}{$1$}
\psfrag{E}{e}
\includegraphics[width=.95\linewidth]{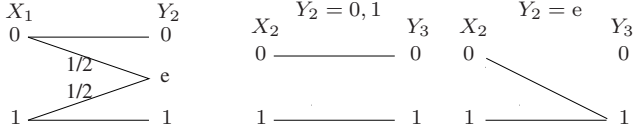}
\caption{Channel statistics of the degraded noncausal relay channel}
\label{DFGPvsDF}
\end{figure}

Note that the state of the channel from the relay to the receiver, namely,
whether we get an erasure or not, is independent of $X_1$.
The first term in both the DF lower bound and the GP-DF lower bound is easy to
compute as
\[
\max_{p(x_1)} I(X_1;Y_2) = 1/2.
\]
Consider the second term in the DF lower bound. Here $X_2$ is chosen such that
$Y_2 \rightarrow X_1 \rightarrow X_2$ form a Markov chain. By carefully
computing the conditional probability $p(y_3|x_1,x_2) = \sum_{y_2}
p(y_2|x_1)p(y_3|x_2,y_2)$ in this specific channel, we can show that $X_1
\rightarrow X_2 \rightarrow Y_3$ form a Markov chain. Thus,
\begin{align*}
\max_{p(x_1,x_2)} I(X_1,X_2;Y_3) &\stackrel{(a)}{=} \max_{p(x_2|x_1)}
I(X_2;Y_3)\\
& \stackrel{(b)}{=} \max_{p(x_2)} I(X_2;Y_3)\\
& \stackrel{(c)}{=} H(1/5)-2/5\\
& = 0.3219,
\end{align*}
where $(a)$ follows since $X_1 \rightarrow X_2 \rightarrow Y_3$ form a Markov
chain, $(b)$ follows since $I(X_2;Y_3)$ is fully determined by the marginal
distribution $p(x_2,y_3)$, and $(c)$ follows since the channel from $X_2$ to
$Y_3$ $p(y_3|x_2) = \sum_{y_2}p(y_3|x_2,y_2)p(y_2)$ is a $Z$ channel with
crossover probability 1/2 regardless of $p(x_1)$. Thus,
\[
R_{\textrm{DF}} = \min\{1/2,\; 0.3219\} = 0.3219.
\]
Now consider the second term in the GP-DF lower bound~(\ref{DFGP})
\[
\max_{\substack{p(x_1)p(u_2|x_1,y_2)\\x_2(u,x_1,x_2,y_2)}}
[I(X_1,U;Y_3)-I(U;Y_2|X_1)].
\]
Let $U = X_2 = 1$, if $Y_2 = \mathrm{e}$, and $U=X_2 = \mathrm{Bern}(1/2)$, if
$Y_2 = 0,1$. Note that here we always have $Y_3 = X_2 = U$ and $X_1 \rightarrow
Y_2 \rightarrow X_2$ form a Markov chain. Thus,
\begin{align*}
&{I(X_1,U;Y_3)-I(U;Y_2|X_1)}\\
&= I(X_1,X_2;X_2)-I(X_2;Y_2|X_1)\\
& = H(X_2) - H(X_2|X_1) + H(X_2|Y_2,X_1)\\
& = I(X_1;X_2) + H(X_2|Y_2)\\
& \geq H(X_2|Y_2) \\
& = 1/2.
\end{align*}
Therefore,
\[
R_{\textrm{GP-DF}} = 1/2 > R_{\textrm{DF}} = 0.3219.
\]
Moreover, it is easy to see from the cutset bound~(\ref{CutUpBnd}) that the rate
1/2 is also an upper bound and hence $C_{\infty} = 1/2$.
\end{example}


\subsection{Gelfand--Pinsker Compress--Forward Lower Bound}
\label{sec:CF}
In this subsection, we first propose a two-stage coding scheme that incorporates
Gelfand--Pinsker coding with the compress--forward coding scheme. Then we show
an equivalent lower bound can be established directly by applying the recently
developed hybrid coding scheme at the relay node.


\begin{theorem}[Gelfand--Pinsker compress--forward (GP-CF) lower bound]
\label{thm:GPCF1}
The capacity of the noncausal relay channel is lower bounded as
\begin{align}
&C_{\infty} \geq R_{\emph{GP-CF}}\nonumber\\
&= \max \min\{I(X_1;U,Y_3),\;
I(X_1,U;Y_3)-I(U;Y_2|X_1)\},
\label{eq:GPCF2}
\end{align}
where the maximum is over all pmfs $p(x_1)p(u|y_2)$ and functions $x_2(u,y_2)$.
\end{theorem}

\begin{IEEEproof}[Outline of the proof]
The coding scheme is illustrated in Figure \ref{CF_GP_bin}. We use Wyner--Ziv
binning, multicoding, and joint typicality encoding and decoding. A description
$\hat{y}_2^n$ of $y_2^n$ is constructed at the relay. Since the receiver has
side information $y_3^n$ about $\hat{y}_2^n$, we use binning as in Wyner--Ziv
coding to reduce the rate necessary to send $\hat{y}_2^n$. Since the relay has
side information $y_2^n$ of the channel $p(y_3|x_1,x_2,y_2)$, we use multicoding
as in Gelfand--Pinsker coding to  send the bin index of $\hat{y}_2^n$ via $u^n$.
The decoder first decode the bin index from $u^n$. It then uses $u^n$ and
$y_3^n$ to decode $\hat{y}_2^n$ and $x_1^n(m)$ simultaneously.

\begin{figure}[htbp]
\centering
\small
\psfrag{k}{$\hat{y}_2^n(k)$}
\psfrag{lm}{$l_m$}
\psfrag{l}{$u^n(l|l_m)$}
\psfrag{hatR2}{${\Large 2^{n\hat{R}_2}}$}
\psfrag{tildeR2}{${\Large 2^{n\tilde{R}_2}}$}
\psfrag{B1}{$\mathcal{B}(1)$}
\psfrag{B2}{$\mathcal{B}(2)$}
\psfrag{Bl}{$\mathcal{B}(2^{nR_2})$}
\psfrag{C1}{$\mathcal{C}(1)$}
\psfrag{C2}{$\mathcal{C}(2)$}
\psfrag{Cl}{$\mathcal{C}(2^{nR_2})$}
\psfrag{1}{$1$}
\psfrag{2}{$2$}
\psfrag{R2}{$2^{nR_2}$}
\psfrag{dot1}{${\Large\vdots}$}
\psfrag{dot2}{${\Large\vdots}$}
\psfrag{dot3}{${\Large\vdots}$}
\includegraphics[width=1\linewidth]{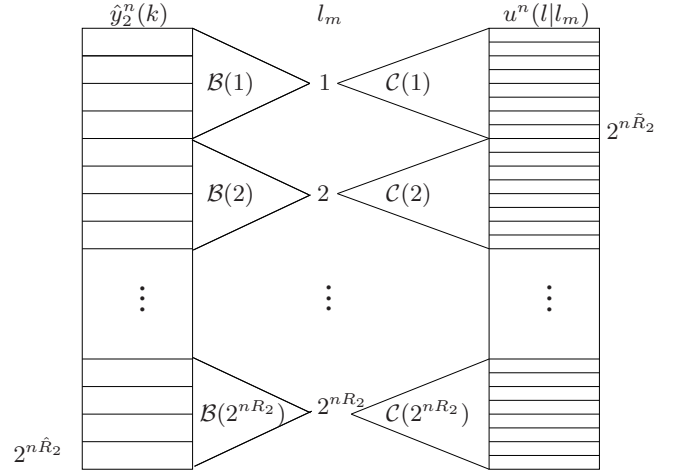}
\caption{GP-CF coding scheme with binning and multicoding}
\label{CF_GP_bin}
\centering
\end{figure}
We now provide the details of the coding scheme.

\smallskip
{\it Codebook generation:} Fix 
$p(x_1)p(u|y_2)p(\hat{y}_2|y_2)$ and $x_2(u,\hat{y_2},y_2)$ that attain the lower
bound. Randomly and independently
 generate $2^{nR}$ sequences $x_1^n(m)$, $m\in [1:2^{nR}]$, each according to
$\prod_{i=1}^n P_{X_1}(x_{1i})$.
 Randomly and independently generate $2^{n\hat{R}_2}$ sequences
$\hat{y}_2^n(k)$, $k\in [1:2^{n\hat{R}_2}]$,
 each according to $\prod_{i=1}^n P_{\hat{Y}_2}(\hat{y}_{2i})$. Partition $k$
into $2^{nR_2}$ bins $\mathcal{B}(l_m)$.
 For each $l_m$, randomly and independently generate $2^{n\tilde{R}_2}$
sequences $u^n(l|l_m)$, $l \in [1:2^{n\tilde{R}_2}]$,
 each according to $\prod_{i=1}^n P_U(u_i)$, which form subcodebook
$\mathcal{C}(l_m)$. This defines the codebook
 $\mathcal{C} = \{(x_1^n(m), \hat{y}_2^n(k), u^n(l|l_m),x_2^n(u^n,\hat{y}_2^n,
y_2^n)): m\in[1:2^{nR}], k\in[1:2^{n\hat{R}_2}],
  l_m \in [1:2^{nR_2}], l \in [1:2^{n\tilde{R}_2}]\}$. The codebook is revealed
to all parties. 

\smallskip
{\it Encoding:} To send the message $m$, the encoder transmits
$x_1^n(m)$.

\smallskip
{\it Relay encoding and analysis of the probability of error:} Upon receiving
$y_2^n$, the relay first finds the unique $k$ such that
$(\hat{y}_2^n(k),y_2^n)\in
\mathcal{T}_{\epsilon'}^{(n)}$. This requires $\hat{R}_2 >
I(\hat{Y}_2;Y_2)+\delta(\epsilon')$ by the covering lemma. Upon getting the bin
index $l_m$ of $k$, i.e., $k \in \mathcal{B}(l_m)$, the relay finds a sequence
$u^n(l|l_m)\in \mathcal{C}(l_m)$ such that $(u^n(l|l_m),y_2^n)\in
\mathcal{T}_{\epsilon'}^{(n)}$. This requires $\tilde{R}_2 >
I(U;Y_2)+\delta(\epsilon')$ by the covering lemma.
The relay transmits $x_2(\hat{y}_{2i}(k),u_i(l|l_m),y_{2i})$ at time $i \in
[1:n]$.

\smallskip
{\it Decoding and analysis of the probability of error:} Let $\epsilon >
\epsilon'$. Upon receiving $y_3^n$, the decoder finds the unique $\hat{l}_m$
such that
$(u^n(l|\lh_m),y_3^n) \in \mathcal{T}_{\epsilon}^{(n)}$ for some
$u^n(l|\lh_m)\in \mathcal{C}(\lh_m)$. This requires $\tilde{R}_2+R_2 <
I(U;Y_3)-\delta(\epsilon)$.
The decoder then finds the unique message $\mh$ such that $(x_1^n(\mh),
\hat{y}_2^n(k), y_3^n) \in \aep$ for some
$k \in \mathcal{B}(\lh_m)$. Let $K$ be the chosen index for $\hat{Y}_2^n$ at the
relay. If $\mh \neq 1$ but $k = K$, this requires $R< I(X_1;\hat{Y}_2,
Y_3)-\delta(\epsilon)$. If $\mh \neq 1$ and $k \neq K$, this requires
$R+\hat{R}_2-R_2 < I(X_1;Y_3)+I(\hat{Y}_2;X_1,Y_3)-\delta(\epsilon)$. Thus, we
establish the following lower bound:
\begin{align}
C_{\infty} &\geq R'_{\textrm{GP-CF}} \nonumber\\
&= \max \min\{I(X_1;\hat{Y}_2,Y_3),\nonumber\\
&I(X_1,\Yh_2;Y_3)-I(\hat{Y}_2;Y_2|X_1)+I(U;Y_3)-I(U;Y_2)\},
\label{eq:CFGP1}
\end{align}
where the maximum is over all pmfs $p(x_1)p(u|y_2)p(\hat{y}_2|y_2)$ and
functions $x_2(u,\hat{y}_2,y_2)$.

Now we show the two lower bounds (\ref{eq:CFGP1}) and (\ref{eq:GPCF2}) are
equivalent.
Setting $U = \emptyset$ in $R'_{\textrm{GP-CF}}$ and relabeling $\Yh_2$ as $U$,
$R'_{\textrm{GP-CF}}$ reduces to $R_{\textrm{GP-CF}}$. Thus,
\begin{equation}
R'_{\textrm{GP-CF}} \geq R_{\textrm{GP-CF}}.
\label{eq:eqv1}
\end{equation}
On the other hand, letting $U = (U,\hat{Y}_2)$ in $R_{\textrm{GP-CF}}$, we have
\begin{align*}
& I(X_1,U,\Yh_2;Y_3) - I(U,\Yh_2;Y_2|X_1)\\
&= I(X_1,\Yh_2;Y_3) + I(U;Y_3|X_1,\Yh_2)\\
&\hspace{1em}-I(\Yh_2;Y_2|X_1)-I(U;Y_2|X_1,\Yh_2)\\
&= I(X_1,\Yh_2;Y_3)- I(\Yh_2;Y_2|X_1) + H(U|X_1,\Yh_2) \\
&\hspace{1em}-H(U|X_1,\Yh_2,Y_3) - H(U|X_1,\Yh_2)+H(U|X_1,\Yh_2,Y_2) \\
&\stackrel{(a)}{\geq} I(X_1,\Yh_2;Y_3)- I(\Yh_2;Y_2|X_1) + H(U)\\
&\hspace{1em}-H(U|Y_3) - H(U) +H(U|X_1,\Yh_2,Y_2)\\
&\stackrel{(b)}{=} I(X_1,\Yh_2;Y_3)- I(\Yh_2;Y_2|X_1) + I(U;Y_3) -
I(U;Y_2),
\end{align*}
where $(a)$ follows since conditioning reduces entropy and $(b)$ follows since
$(X_1,\Yh_2) \rightarrow Y_2 \rightarrow U$ form a Markov chain. Furthermore,
since the maximum in $R_{\textrm{GP-CF}}$ is over a larger set
$p(u,\hat{y}_2|y_2)$ than the set $p(u|y_2)p(\hat{y}_2|y_2)$ in
$R'_{\textrm{GP-CF}}$,
\begin{equation}
R_{\textrm{GP-CF}} \geq R'_{\textrm{GP-CF}}.
\label{eq:eqv2}
\end{equation}
Combining (\ref{eq:eqv1}) and (\ref{eq:eqv2}) establishes the equivalence.
\end{IEEEproof}

\begin{remarks}
Taking $U$ independent of $Y_2$ and $X_2 = U$ in (\ref{eq:CFGP1}), we establish
the compress--forward lower bound without Gelfand--Pinsker coding as follows:
\begin{align*}
C_{\infty} &\geq R_{\mathrm{CF}}\\
&= \max \min \{ I(X_1;\hat{Y}_2,Y_3),\\
&I(X_1,\Yh_2;Y_3) + I(X_2;Y_3)-I(\hat{Y}_2;Y_2|X_1)\},
\end{align*}
where the maximum is over all pmfs $p(x_1)p(x_2)p(\hat{y}_2|y_2)$.
\end{remarks}

In the analysis of the probability of error in Theorem~\ref{thm:GPCF1}, there is
a technical subtlety in applying the standard packing lemma and joint typicality
lemma, since the bin index $L_m$, the compression index $K$, and the multicoding
index $L$ all depend on the random codebook itself. In the following, we show
the GP-CF lower bound (\ref{eq:GPCF2}) can be established directly by applying
the recently developed hybrid coding scheme for joint source--channel coding by
Lim, Minero, and Kim \cite{Lim2010}, \cite{Lim2011}, \cite{Kim2011}.

\begin{IEEEproof}[Proof of Theorem~\ref{thm:GPCF1} via hybrid coding]
\begin{figure}[htbp]
\centering
\small
\psfrag{s}[b]{$Y_2^n$}
\psfrag{e1}[cc]{Vector}
\psfrag{e}[cc]{quantizer}
\psfrag{m}[b]{$U^n(L)$}
\psfrag{e3}[cc]{$x_2(u,y_2)$}
\psfrag{x}[b]{$X_2^n$}
\includegraphics[width = .9\linewidth]{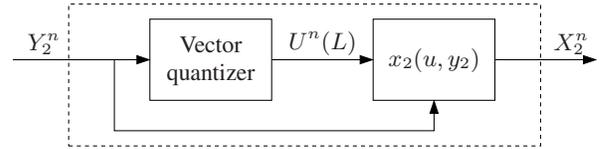}
\caption{Hybrid coding interface at the relay. Illustration from Kim, Lim, and
Minero \cite{Kim2011}}
\label{fig:hybrid}
\end{figure}
In this coding scheme, we apply hybrid coding at the relay node as depicted in
Figure~\ref{fig:hybrid}. The sequence $y_2^n$ is mapped to one of $2^{n\Rt}$
sequences $u^n(l)$. The relay generates the codeword $x_2^n$ through a {\it
symbol-by-symbol} mapping $x_2(u,y_2)$. The receiver declares $\mh$ to be the
message estimate if $(x_1^n(\mh),u^n(l),y_3^n)$ are jointly typical for some
$l\in [1::2^{n\Rt}]$. Similar to the hybrid coding scheme for joint
source-channel coding \cite{Lim2010}, \cite{Lim2011}, \cite{Kim2011}, the
precise analysis
 of the probability of decoding error involves a technical subtlety. In
particular, since $U^n(L)$ is used as a source codeword,
 the index $L$ depends on the entire codebook. This dependency issue is resolved
by the technique developed in \cite{Lim2010}.
 We now provide the details of the coding scheme. 
 
 \smallskip
{\it Codebook generation:} Fix $p(x_1)p(u|y_2)$ and $x_2(u,y_2)$ that attain the
lower bound. Randomly and independently generate
$2^{nR}$ sequences $x_1^n(m)$, $m \in [1:2^{nR}]$, each according to
$\prod_{i=1}^n p_{X_1}(x_{1i})$. Randomly and independently generate
$2^{n\Rt}$ sequences $u^n(l)$, $l \in [1:2^{n\Rt}]$, each according to
$\prod_{i=1}^n p_{U}(u_{i})$.
This defines the codebook $\mathcal{C}=\{(x_1^n(m), u^n(l),
x_2^n(u^n(l),y_2^n)): m \in [1:2^{nR}], l\in [1:2^{n\Rt}]\}$.
The codebook is revealed to all parties. 

\smallskip
{\it Encoding:} To send message $m$, the encoder transmits
$x_1^n(m)$.

\smallskip
{\it Relay encoding:} Upon receiving $y_2^n$, the relay finds an index $l$ such
that $(u^n(l),y_2^n) \in \mathcal{T}_{\epsilon'}^{(n)}$. If there is more than
one such indices, it chooses one of them at random. If there is no such index,
it chooses an arbitrary index at random from $[1::2^{n\Rt}]$.
The relay then transmits $x_{2i}(u_{i}(l),y_{2i})$ at time $i \in
[1:n]$.

\smallskip
{\it Decoding:} Let $\epsilon > \epsilon'$. Upon receiving $y_3^n$, the decoder
finds the unique message $\mh$ such that
$(x_1^n(\mh), u^n(l), y_3^n) \in \mathcal{T}_{\epsilon}^{(n)}$ for some $l \in
[1::2^{n\Rt}]$.

\smallskip
{\it Analysis of the probability of error:}
We analyze the probability of decoding error averaged over codes. Let $L$ denote
the index of the chosen $U^n$ codeword for $Y_2^n$. Assume without loss of
generality that $M = 1$. The decoder makes an error only if one of the following
events occur:
\begin{align*}
\tilde{\mathcal{E}} &= \{(U^n(l),Y_2^n) \notin \mathcal{T}_{\epsilon'}^{(n)}\;
\textrm{for all}\; l\},\\
\mathcal{E}_1 &= \{(X_1^n(1), U^n(L),Y_3^n) \notin
\mathcal{T}_{\epsilon}^{(n)}\},\\
\mathcal{E}_2 &= \{(X_1^n(m),U^n(L),Y_3^n) \in \mathcal{T}_{\epsilon}^{(n)}\;
\textrm{for}\; m\neq 1\},\\
\mathcal{E}_3 &= \{(X_1^n(m),U^n(l),Y_3^n) \in \mathcal{T}_{\epsilon}^{(n)}\;
\textrm{for}\; m\neq 1, l \neq L\}.
\end{align*}

By the union of the events bound, the probability of error is upper bounded as
\begin{align*}
\P(\mathcal{E}) &= \P(\tilde{\mathcal{E}} \cup \mathcal{E}_1 \cup \mathcal{E}_2
\cup \mathcal{E}_3)\\
&\leq \P(\tilde{\mathcal{E}})+ \P(\mathcal{E}_1 \cap \tilde{\mathcal{E}}^c) +
\P(\mathcal{E}_2 \cap \tilde{\mathcal{E}}^c) + \P(\mathcal{E}_3).
\end{align*}
By the covering lemma, the first term tends to zero as $n \rightarrow \infty$ if
$\Rt>I(U;Y_2) + \delta(\epsilon')$.
By the conditional typicality lemma, the second term tends to zero as $n
\rightarrow \infty$. By the packing lemma, the third term tends to zero as $n
\rightarrow \infty$ if $R < I(X_1;U,Y_3)-\delta(\epsilon)$.

The fourth term requires special attention. Consider
\begin{align*}
\P(\Ec_3) &= \P\{(X_1^n(m),U^n(l),Y_3^n) \in \mathcal{T}_{\epsilon}^{(n)}\;
\textrm{for}\; m\neq 1, l \neq L\} \\
& \stackrel{(a)}{\leq} \sum_{m=2}^{2^{nR}} \sum_{l = 1}^{2^{n\Rt}}
\P\{(X_1^n(m),U^n(l),Y_3^n) \in \mathcal{T}_{\epsilon}^{(n)}, l \neq L\} \\
& = \sum_{m=2}^{2^{nR}} \sum_{l = 1}^{2^{n\Rt}} \sum_{y_2^n} p(y_2^n)\\
&\hspace{1em}\cdot\P\{(X_1^n(m),U^n(l),Y_3^n) \in \mathcal{T}_{\epsilon}^{(n)}, l \neq L |Y_2^n = y_2^n\}  \\
& \stackrel{(b)}{\leq} 2^{nR} 2^{n\Rt} \sum_{y_2^n}p(y_2^n)\\
&\hspace{1em}\cdot\P\{(X_1^n(2),U^n(1),Y_3^n)
\in \mathcal{T}_{\epsilon}^{(n)}, L \neq 1 | Y_2^n = y_2^n\},
\end{align*}
where $(a)$ follows by the union of events bound and $(b)$ follows by the
symmetry of the codebook generation and relay encoding. Let $\bar{\Cc} = \Cc
\setminus \{(X_1^n(2),U^n(1), X_2^n(X_1^n(2),U^n(1)))\}$. Then, for $n$
sufficiently large,
\begin{align}
& \P\{(X_1^n(2),U^n(1),Y_3^n) \in \aep, L \neq 1 | Y_2^n = y_2^n\} \nonumber\\
& \leq \P\{(X_1^n(2),U^n(1),Y_3^n) \in \aep | L \neq 1, Y_2^n = y_2^n\}
\nonumber\\
& = \sum_{(x_1^n,u^n,y_3^n) \in \aep} \P\{X_1^n(2) = x_1^n,U^n(1) = u^n,Y_3^n =
y_3^n \nonumber\\[-1em]
&\hspace{17em}|L \neq 1, Y_2^n = y_2^n\} \nonumber\\
& = \sum_{(x_1^n,u^n,y_3^n) \in \aep} \sum_{\bar{\Cc}} \P\{\bar{\Cc} = \bar{\cc}| L \neq 1, Y_2^n = y_2^n\} \nonumber\\
&\hspace{1em}\cdot\P\{X_1^n(2) = x_1^n,U^n(1) = u^n,Y_3^n = y_3^n\nonumber\\
&\hspace{14em}| L \neq 1, Y_2^n = y_2^n, \bar{\Cc} =\bar{\cc}\} \nonumber\\
& \stackrel{(a)}{=} \sum_{(x_1^n,u^n,y_3^n) \in \aep} \sum_{\bar{\Cc}}
\P\{\bar{\Cc} = \bar{\cc}| L \neq 1, Y_2^n = y_2^n\}
\nonumber\\
&\hspace{1em}\cdot\P\{U^n(1) = u^n| L \neq 1, Y_2^n = y_2^n, \bar{\Cc} = \bar{\cc}\}\nonumber\\
&\hspace{1em}\cdot\P\{X_1^n(2)= x_1^n| L \neq 1, Y_2^n = y_2^n,\bar{\Cc} = \bar{\cc},Y_3^n = y_3^n\}\nonumber\\
&\hspace{1em} \cdot \P\{Y_3^n = y_3^n| L \neq 1, Y_2^n = y_2^n, \bar{\Cc} = \bar{\cc}\}  \nonumber\\
& \stackrel{(b)}{\leq} \sum_{(x_1^n,u^n,y_3^n) \in \aep} \sum_{\bar{\Cc}} \P\{\bar{\Cc} = \bar{\cc}| L \neq 1, Y_2^n =
y_2^n\}\nonumber\\
&\hspace{1em}\cdot 2\P\{U^n(1) = u^n\} \P\{X_1^n(2) = x_1^n\}\nonumber\\
&\hspace{1em}\cdot \P\{Y_3^n = y_3^n| L \neq 1, Y_2^n =
y_2^n, \bar{\Cc} = \bar{\cc}\} \nonumber\\
& = \sum_{(x_1^n,u^n,y_3^n) \in \aep} 2\P\{U^n(1) = u^n\} \P\{X_1^n(2) = x_1^n\}\nonumber\\
&\hspace{1em}\cdot\P\{Y_3^n = y_3^n| L \neq 1, Y_2^n = y_2^n\} \nonumber\\
& \stackrel{(c)}{\leq} \sum_{(x_1^n,u^n,y_3^n) \in \aep} 2\P\{U^n(1) =u^n\}\P\{X_1^n(2) = x_1^n\}\nonumber\\
&\hspace{1em}\cdot 2\P\{Y_3^n = y_3^n|Y_2^n = y_2^n\},
\label{eq:pe3}
\end{align}
where $(a)$ follows since given $L \neq 1$, $U^n(1) \rightarrow
(Y_2^n,\bar{\Cc}) \rightarrow (Y_3^n,X_1^n(2))$ form a Markov chain, $(b)$
follows since for $n$ sufficiently large $\P\{U^n(1) = u^n| L \neq 1, Y_2^n =
y_2^n, \bar{\Cc} = \bar{\cc}\} \leq 2\P\{U^n(1) = u^n\}$ and $X_1^n(2)$ is
independent of $(Y_2^n,Y_3^n, \bar{\Cc}, K)$, and $(c)$ follows since for $n$
sufficiently large $\P\{Y_3^n = y_3^n| L \neq 1, Y_2^n = y_2^n\} \leq 2\P\{Y_3^n
= y_3^n|Y_2^n = y_2^n\}$. The statements in $(b)$ and $(c)$ are established by
Lim, Minero, and Kim in \cite[Lemmas 1, 2]{Lim2010}. Back to the upper bound on
$\P(\Ec_3)$, by the joint typicality lemma and (\ref{eq:pe3}), we have
\begin{align*}
&\P(\Ec_3) \\
&= \P\{(X_1^n(m),U^n(l),Y_3^n) \in \mathcal{T}_{\epsilon}^{(n)}\;
\textrm{for}\; m\neq 1, l \neq L\} \\
& \leq 4 \cdot 2^{nR} 2^{n\Rt} \sum_{y_2^n}p(y_2^n) \sum_{(x_1^n,u^n,y_3^n) \in
\aep} p(y_3^n|y_2^n)\\
&\hspace{5em}\cdot\P\{X_1^n(2) = x_1^n\}\P\{U^n(1) = u^n\}  \\
& = 4 \cdot 2^{nR} 2^{n\Rt} \sum_{(x_1^n,u^n,y_3^n) \in \aep} p(y_3^n)\\
&\hspace{5em}\cdot\P\{X_1^n(2) =
x_1^n\}\P\{U^n(1) = u^n\} \\
& \leq 4 \cdot 2^{n(R+\Rt-I(X_1;Y_3)-I(U;X_1,Y_3)+\d(\e))},
\end{align*}
which tends to zero as $n \rightarrow \infty$ if $R+\Rt \leq
I(X_1;Y_3)+I(U;X_1,Y_3)-\d(\e)$. Eliminating $\Rt$ and letting $n \rightarrow
\infty$ completes the proof.
\end{IEEEproof}

\subsection{Gelfand--Pinsker Partial Decode--Forward Compress--Forward Lower
Bound}
\label{sec:DFCF}
Finally, we further combine the hybrid coding scheme developed for the GP-CF
lower bound with the partial decode--forward coding scheme by El~Gamal,
Hassanpour, and Mammen~\cite{Hassanpour07}.

\begin{theorem}[GP-PDF-CF lower bound]
\label{thm:GPPDFCF}
The capacity of noncausal relay channel is lower bounded as
\begin{align}
&C_{\infty} \geq R_{\textrm{\em GP-PDF-CF}} \nonumber\\
&= \max \min \{I(V,U;Y_3)+I(X_1;U,Y_3|V)-I(U;Y_2|V),\nonumber\\
&\hspace{5.5em} I(V;Y_2)+I(X_1;U,Y_3|V), \nonumber\\
&\hspace{5.5em} I(V;Y_2)+I(X_1;U,Y_3|V)\nonumber\\
&\hspace{7em}+I(U;Y_3|V)-I(U;Y_2|V)\},
\label{PDFCFGP}
\end{align}
where the maximum is over all pmfs $p(v,x_1) p(u|v,y_2)$ and functions
$x_2(u,v,y_2)$.
\end{theorem}

\begin{remarks}
Setting $V=(V,X_2)$ and $U=\emptyset$ reduces the GP-PDF-CF lower bound to the
PDF lower bound (\ref{PDecLwBnd}). Note that such choice induces the Markov
chains $X_2 \rightarrow V \rightarrow Y_2$ and $V \rightarrow (X_1,X_2)
\rightarrow Y_3$. Furthermore, setting $V=\emptyset$ reduces the GP-PDF-CF lower
bound to the GP-CF lower bound (\ref{eq:GPCF2}). The fact that the GP-PDF-CF
lower bound recovers the GP-DF lower bound (\ref{DFGP}) requires a special care.
See Appendix~\ref{AppendixA} for a complete proof.
\end{remarks}

\begin{IEEEproof}
In this coding scheme, message $m \in [1:2^{nR}]$ is divided into two
independent parts $m'$ and $m''$ where $m' \in [1:2^{nR'}]$, $m'' \in
[1:2^{nR''}]$, and $R'+R''=R$. For each message $m=(m',m'')$, we generate a
$x_1^n(m''|m')$ sequence and a {\it subcodebook} $\mathcal{C}(m')$ of
$2^{n\tilde{R}}$ $u^n(l|m')$ sequences. To send message $m=(m',m'')$, the sender
transmits $x_1^n(m''|m')$. Upon receiving $y_2^n$ noncausally, the relay
recovers the message $\tilde{m}'$, finds a $u^n(l|\tilde{m}') \in
\mathcal{C}(\tilde{m}')$ that is jointly typical with $(v^n(\tilde{m}'),y_2^n)$,
and transmits $x_2^n(u^n(l|\tilde{m}'),v^n(\tilde{m}'),y_2^n)$, as illustrated
in Figure~\ref{fig:GPPDFCF}. The receiver declares
$\hat{m}=(\hat{m}',\hat{m}'')$ to be the message estimate if
$(x_1^n(\hat{m}''|\hat{m}'),u^n(l|\hat{m}'),v^n(\hat{m}'),y_3^n)$ are jointly
typical for some $u^n(l|\hat{m}')\in \mathcal{C}(\hat{m}')$.
\begin{figure}[htbp]
\centering
\small
\psfrag{Y2}[l]{$Y_2^n$}
\psfrag{rec}[cc]{Recover}
\psfrag{Mp}[cc]{$M'$}
\psfrag{vector}[cc]{\hspace{.6em}Vector}
\psfrag{quantizer}[cc]{\hspace{.6em}quantizer}
\psfrag{U}[bc]{$U^n(L|M')$}
\psfrag{V}[bc]{$V^n(M')$}
\psfrag{X2fun}[cc]{$\hspace{.4em}x_2(v,u,y_2)$}
\psfrag{X2}[l]{$X_2^n$}
\includegraphics[width=.9\linewidth]{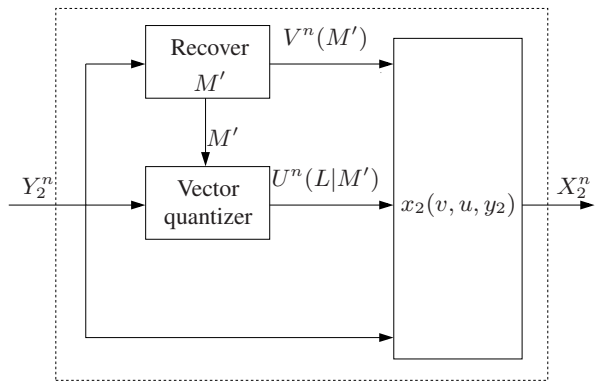}
\caption{Hybrid coding interface at the relay.}
\label{fig:GPPDFCF}
\end{figure}

We now provide the details of the coding scheme. 

\smallskip
{\it Codebook generation:} Fix $p(v,x_1) p(u|v,y_2) x_2(u,v,y_2)$ that attains
the lower bound. Randomly and independently generate $2^{nR'}$ sequences
$v^n(m')$, $m' \in [1:2^{n R'}]$, each according to $\prod_{i=1}^n
p_{V}(v_{i})$.
For each message $m' \in [1:2^{nR'}]$, randomly and conditionally independently
generate $2^{nR''}$ sequences $x_1^n(m''|m')$
and $2^{n\tilde{R}}$ sequences $u^n(l|m')$, each respectively according to
$\prod_{i=1}^n p_{X_1|V}(x_{1i}|v_{i}(m'))$ and
$\prod_{i=1}^n p_{U|V}(u_{i}|v_{i}(m'))$, which form the subcodebook
$\mathcal{C}(m')$. This defines the codebook
$\mathcal{C}=\{(v^n(m'),x_1^n(m''|m'),u^n(l|m'),x_2^n(u^n(l|m'),v^n(m'),y_2^n))
: m'\in[1:2^{nR'}], m''\in[1:2^{nR''}], l\in[1:2^{n\tilde{R}}]\}$.
 The codebook is revealed to all parties.
 
\smallskip
{\it Encoding:} To send message $m=(m',m'')$, the encoder transmits
$x_1^n(m''|m')$.

\smallskip
{\it Relay encoding:} Upon receiving $y_2^n$, the relay finds the unique
$\tilde{m}'$ such that $(v^n(\tilde{m}'),y_2^n) \in
\mathcal{T}_{\epsilon'}^{(n)}$. Then, it finds the unique sequence
$u^n(l|\tilde{m}') \in \mathcal{C}(\tilde{m}')$
such that $(u^n(l|\tilde{m}'),v^n(\tilde{m}'),y_2^n) \in
\mathcal{T}_{\epsilon'}^{(n)}$. If there is more than one such index, it chooses
one of them at random. If there is no such index, it chooses an arbitrary index
at random from $[1::2^{n\Rt}]$.
The relay then transmits $x_{2i} =
x_{2}(u_i(l|\tilde{m}'),v_i(\tilde{m}'),y_{2i})$ at time $i \in
[1:n]$.

\smallskip
{\it Decoding:} Let $\epsilon > \epsilon'$. Upon receiving $y_3^n$, the decoder
declares that $\hat{m} = (\hat{m}',\hat{m}'') \in [1:2^{nR}]$ is sent if it is
the unique message such that
$(x_1^n(\hat{m}''|\hat{m}'),u^n(l|\hat{m}'),v^n(\hat{m}'),
y_3^n) \in \mathcal{T}_{\epsilon}^{(n)}$ for some $u^n(l|\hat{m}')\in
\mathcal{C}(\hat{m}')$; otherwise, it declares an error.

\smallskip
{\it Analysis of the probability of error:} We analyze the probability of error
of message $M$ averaged over codes.
Assume without loss of generality that $M=(M',M'')=(1,1)$. Let $\tilde{M}'$ be
the decoded message at the relay and let $L$ denote the index of the chosen
$U^n$ codeword for $\tilde{M}'$. The decoder makes an error only if one of the
following events occur:
\begin{align*}
\tilde{\mathcal{E}} &= \{\tilde{M}' \neq 1\},\\
\tilde{\mathcal{E}}_1 &= \{(V^n(1),Y_2^n) \notin
\mathcal{T}_{\epsilon'}^{(n)}\},\\
\tilde{\mathcal{E}}_2 &= \{(V^n(m'),Y_2^n) \in \mathcal{T}_{\epsilon'}^{(n)}
\;\textrm{for some}\; m' \neq 1\},\\
\tilde{\mathcal{E}}_3 &= \{(U^n(l|\tilde{M}'), V^n(\tilde{M}'), Y_2^n) \notin
\mathcal{T}_{\epsilon'}^{(n)} \\
&\hspace{11em}\textrm{for all}\; U^n(l|\tilde{M}')\in\mathcal{C}(\tilde{M}')\},\\
\mathcal{E}_1 &= \{(X_1^n(1|\tilde{M}'),U^n(L|\tilde{M}'),V^n(\tilde{M}'),Y_3^n)
\notin \mathcal{T}_{\epsilon}^{(n)}\},\\
\mathcal{E}_2 &= \{(X_1^n(m''|1),U^n(L|1),V^n(1),Y_3^n) \in
\mathcal{T}_{\epsilon}^{(n)} \\
&\hspace{15em}\textrm{for some}\; m''\neq 1\},\\
\mathcal{E}_3 &= \{(X_1^n(m''|1),U^n(l|1),V^n(1),Y_3^n) \in
\mathcal{T}_{\epsilon}^{(n)} \\
&\hspace{3.5em}\textrm{for some}\; m''\neq 1, l \neq L,
\;\textrm{and}\; U^n{(l|1)\in \mathcal{C}(1)}\},\\
\mathcal{E}_4 &= \{(X_1^n(m''|m'),U^n(l|m'),V^n(m'),Y_3^n) \in
\mathcal{T}_{\epsilon}^{(n)} \\
&\hspace{5.5em}\textrm{for some}\; m'\neq 1, m'', U^n{(l|m')\in
\mathcal{C}(m')}\}.
\end{align*}
By the union of events bound, the probability of error is upper bounded as
\begin{align*}
\P(\mathcal{E}) &= \P(\hat{M}\neq 1)\\
&\leq \P(\tilde{\mathcal{E}} \cup \tilde{\mathcal{E}}_3 \cup \mathcal{E}_1 \cup
\mathcal{E}_2 \cup \mathcal{E}_3 \cup \mathcal{E}_4)\\
&\leq \P(\tilde{\mathcal{E}}) + \P(\tilde{\mathcal{E}}_3 \cap
\tilde{\mathcal{E}}^c) + \P(\mathcal{E}_1 \cap \tilde{\mathcal{E}}^c \cap
\tilde{\mathcal{E}}_3^c) \\
&\hspace{2em}+ \P(\mathcal{E}_2) + \P(\mathcal{E}_3) +
\P(\mathcal{E}_4)\\
&\leq \P(\tilde{\mathcal{E}}_1) + \P(\tilde{\mathcal{E}}_2)+
\P(\tilde{\mathcal{E}}_3 \cap \tilde{\mathcal{E}}^c) + \P(\mathcal{E}_1 \cap
\tilde{\mathcal{E}}^c \cap \tilde{\mathcal{E}}_3^c)\\
&\hspace{2em}+ \P(\mathcal{E}_2) +
\P(\mathcal{E}_3) + \P(\mathcal{E}_4).
\end{align*}
By the LLN, the first term tends to zero as $n \rightarrow \infty$. By the
packing lemma, the second term tends to zero as
$n \rightarrow \infty$ if $R'<I(V;Y_2)-\delta(\epsilon')$. Therefore,
$\P(\tilde{\mathcal{E}})$ tends to zero as
$n \rightarrow \infty$ if $R'<I(V;Y_2)-\delta(\epsilon')$. Given
$\tilde{\mathcal{E}}^c$, i.e. $\{\tilde{M}=1\}$, by the covering lemma,
the third term tends to zero as $n \rightarrow \infty$ if
$\tilde{R}>I(U;Y_2|V)+\delta(\epsilon')$.
By the conditional typicality lemma, the fourth term tends to zero as $n
\rightarrow \infty$.
By the joint typicality lemma, the fifth term tends to zero as $n \rightarrow
\infty$ if
$R''<I(X_1;U,Y_3|V) - \delta(\epsilon)$.
The last two terms require special attention because of the dependency between
the index $L$ and the codebook
$\mathcal{C}=\{(V^n(m'),X_1^n(m''|m'),U^n(l|m'),X_2^n(U^n(l|m'),V^n(m'),$ $Y_2^n))
\}$. With a similar argument as in the analysis for $\P(\Ec_3)$ in the proof via
hybrid coding of Theorem~\ref{thm:GPCF1}, we can show the last two terms tend to
zero as $n \rightarrow \infty$ if $R''+\tilde{R}<I(X_1;Y_3|V)+I(U;X_1,Y_3|V) -
\delta(\epsilon)$ and $R' + R'' + \tilde{R}<I(V,X_1,U;Y_3) + I(X_1;U|V) -
\delta(\epsilon)$ respectively.
Eliminating $R'$, $R''$, and $\tilde{R}$ and letting $n \rightarrow \infty$
completes the proof.
\end{IEEEproof}

\section{An Improved Upper Bound}
\label{UpperBound}

In this section, we provide an improved upper bound on the capacity, which is
tight for the class of degraded noncausal relay channels.

\begin{theorem}
The capacity of the noncausal relay channel is upper bounded as
\begin{align}
C_{\infty} & \leq R_{\textrm{\em NUB}} \nonumber\\
&= \max \min
\{I(X_1;Y_2)+I(X_1;Y_3|X_2,Y_2),\;\nonumber\\
&\hspace{5.3em} I(X_1,U;Y_3)-I(Y_2;U|X_1)\},
 \label{eq:newUpBd}
\end{align}
where the maximum is over all pmfs $p(x_1)p(u|x_1,y_2)$ and functions
$x_2(u,y_2)$.
\label{ImpUpBnd}
\end{theorem}

\begin{IEEEproof}
The first term in the upper bound follows from the cutset bound
(\ref{CutUpBnd}). To establish the second bound, identify $U_i =
(M,Y_{2,i+1}^n,Y_3^{i-1})$. Let $Q \sim \mathrm{Unif}[1:n]$ be independent of
$(U^n,X_1^n,Y_2^n,Y_3^n)$ and set $U = (U_{Q},Q)$, $X_1 = X_{1Q}, Y_2 = Y_{2Q}$,
and $Y_3 = Y_{3Q}$. We have
{\allowdisplaybreaks
\begin{align*}
nR & = H(M)\\
 &\stackrel{(a)}{\leq} I(M;Y_3^n) + n\epsilon_n\\
& = \sum_{i=1}^n I(M;Y_{3i}|Y_3^{i-1}) + n\epsilon_n\\
& \leq \sum_{i=1}^n I(M,Y_3^{i-1};Y_{3i}) +n\epsilon_n \\
& = \sum_{i=1}^n [I(M,Y_{2,i+1}^n,Y_3^{i-1};Y_{3i})  \nonumber\\[-3pt]
 &\hspace{5em}-I(Y_{2,i+1}^n;Y_{3i}|Y_3^{i-1},M)] + n\epsilon_n\\[-1pt]
& \stackrel{(b)}{=} \sum_{i=1}^n [I(M,Y_{2,i+1}^n,Y_3^{i-1};Y_{3i})
\nonumber\\[-3pt]
 &\hspace{5em}-I(Y_{2i};Y_3^{i-1}|Y_{2,i+1}^n,M)] + n\epsilon_n\\[-1pt]
& \stackrel{(c)}{=} \sum_{i=1}^n [I(X_{1i},M,Y_{2,i+1}^n,Y_3^{i-1};Y_{3i}) 
\nonumber\\[-3pt]
 &\hspace{5em}-I(Y_{2i};Y_3^{i-1}|Y_{2,i+1}^n,M,X_{1i})]+ n\epsilon_n\\[-1pt]
& \stackrel{(d)}{=} \sum_{i=1}^n [I(X_{1i},M,Y_{2,i+1}^n,Y_3^{i-1};Y_{3i}) 
\nonumber\\[-3pt]
 &\hspace{5em}-I(Y_{2i};M,Y_{2,i+1}^n,Y_3^{i-1}|X_{1i})]+n\epsilon_n\\[-1pt]
& = \sum_{i=1}^n [I(X_{1i},U_i;Y_{3i})-I(Y_{2i};U_i|X_{1i})] +
n\epsilon_n\\[0pt]
& = n[I(X_{1Q},U_Q;Y_{3Q}|Q)-I(Y_{2Q};U_Q|X_{1Q},Q)] +n\epsilon_n\\
& = n[I(X_{1Q},U_Q,Q;Y_{3Q})-I(Y_{2Q};U_Q,Q|X_{1Q})]+n\epsilon_n\\
& = n[I(X_1,U;Y_3)-I(Y_2;U|X_1)]+n\epsilon_n,
\end{align*}}%
where $(a)$ follows by Fano's inequality, $(b)$ follows by Csisz\'{a}r sum
identity, $(c)$ follows since $X_{1i}$ is a function of $M$, and $(d)$ follows
since the channel $p(y_2|x_1)$ is memoryless and thus $(Y_{2,i+1}^n,M)
\rightarrow X_{1i} \rightarrow Y_{2i}$ form a Markov chain.  Note that
from the problem definition, $X_1^n(M)$ is a function of $M$. By our choice of
auxiliary random variable $U_i = (M,Y_{2,i+1}^n,Y_3^{i-1})$, $X_{1i} \to (U_i,
Y_{2i}) \to X_{2i}$ form a Markov chain. Thus, the capacity of the noncausal
relay channel $C_\infty$ is upper bounded as
\begin{align*}
C_{\infty} & \leq R_{\textrm{NUB}}\\
&= \max \min\{I(X_1;Y_2)+I(X_1;Y_3|X_2,Y_2),\;\\
&\hspace{5.5em} I(X_1,U;Y_3)-I(Y_2;U|X_1)\},
\end{align*}
where the maximum is over all pmfs $p(x_1)p(u|x_1,y_2)p(x_2|u,y_2)$. Finally, we
show that it suffices to maximize over $p(x_1)p(u|x_1,y_2)$  and functions
$x_2(u,y_2)$.
For any pmf $p(x_1)p(u|x_1,y_2)p(x_2|u,y_2)$, by the
functional representation lemma \cite[Appendix B]{Kim}, there exists a random
variable $V$ independent of $(U,X_1,Y_2)$ such that $X_2$ is a function of
$(U,Y_2,V)$. Now define $\tilde{U} = (U,V)$. Then
\begin{align*}
C_{\infty} &\leq \max_{\substack{p(x_1)p(u|x_1,y_2)\\p(x_2|u,y_2)}} \min
\{I(X_1;Y_2)+I(X_1;Y_3|X_2,Y_2),\; \\[-4pt]
&\hspace{9em} I(X_1,U;Y_3)-I(Y_2;U|X_1)\}\\[2pt]
&\stackrel{(a)}{\leq} \max_{\substack{p(x_1)p(u|x_1,y_2)\\p(v)x_2(y_2,u,v)}}
\min \{I(X_1;Y_2)+I(X_1;Y_3|X_2,Y_2),\; \\[-10pt]
&\hspace{8em}  I(X_1,U,V;Y_3)-I(Y_2;U,V|X_1)\}\\[4pt]
&\stackrel{(b)}{\leq}
\max_{\substack{p(x_1)p(\tilde{u}|x_1,y_2)\\x_2(y_2,\tilde{u})}} \min
\{I(X_1;Y_2) +I(X_1;Y_3|X_2,Y_2),\;\\[-10pt]
&\hspace{9em} I(X_1,\tilde{U};Y_3)-I(Y_2;\tilde{U}|X_1)\},
\end{align*}
where $(a)$ follows from the independence between $V$ and $(U,X_1,Y_2)$ and
$(b)$ follows by enlarging the input pmf $p(v)$ to a more general pmf
$p(v|u,x_1,y_2)$. Thus, there is no loss of generality in restricting $X_2$ to
be a function of $(U,Y_2)$.
\end{IEEEproof}

\begin{remarks}
This upper bound is obviously tighter than the cutset bound
in~\eqref{CutUpBnd} because of the additional penalty term $I(Y_2;U|X_1)$
in~\eqref{eq:newUpBd}.
\end{remarks}

The improved upper bound coincides with the GP-DF lower bound (\ref{DFGP}) when
the channel is degraded, i.e., $p(y_2|x_1)$ $p(y_3|x_1,x_2,y_2) =
p(y_2|x_1)p(y_3|x_2,y_2)$. Note that the term $I(X_1;Y_2)+I(X_1;Y_3|X_2,Y_2)$ in
the improved upper bound reduces to $I(X_1;Y_2)$, which matches the
corresponding term in the GP-DF lower bound.

\begin{theorem}
\label{thm:capacity}
The capacity of the degraded noncausal relay channel $p(y_2|x_1)p(y_3|x_2,y_2)$
is
\begin{equation}
C_{\infty} = \max \min \{I(X_1;Y_2),\; I(X_1,U;Y_3)-I(Y_2;U|X_1)\},
\label{eq:capacity}
\end{equation}
where the maximum is over all pmfs $p(x_1)p(u|x_1,y_2)$ and functions
$x_2(u,y_2)$.
\label{DgdCapacity}
\end{theorem}
The improved upper bound can be strictly tighter than the cutset bound for the
noncausal relay channel . In the following, we provide an example, motivated by
\cite[Example 2]{Heegard83}, where $R_{\textrm{DF}} < R_{\textrm{GP-DF}} =
C_{\infty} = R_{\textrm{NUB}} < R_{\textrm{CS}}$.

\begin{example}
\label{ex:CSvsNUB}
Consider a degraded noncausal relay channel
$p(y_2|x_1)p(y_3|x_1,x_2,y_2) = p(y_2|x_1)p(y_3|x_2,y_2)$ as depicted in Figure
\ref{ex:BSC}.
The channel from the sender to the relay is $\mathrm{BSC}(p_1)$, while the
channel from the relay to the receiver is $\mathrm{BSC}(p_2)$ if $Y_2 = 0$ and
$\mathrm{BSC}(p_3)$ if $Y_2 = 1$.
\begin{figure}[htbp]
\centering
\footnotesize
\psfrag{0}{0}
\psfrag{1}{1}
\psfrag{p1}{$p_1$}
\psfrag{1-p1}{$1-p_1$}
\psfrag{p2}{$p_2$}
\psfrag{1-p2}{$1-p_2$}
\psfrag{p3}{$p_3$}
\psfrag{1-p3}{$1-p_3$}
\psfrag{X1}{$X_1$}
\psfrag{Y2}{$Y_2$}
\psfrag{S0}{$Y_2 = 0$}
\psfrag{S1}{$Y_2 = 1$}
\psfrag{X2}{$X_2$}
\psfrag{Y3}{$Y_3$}
\includegraphics[width=.95\linewidth]{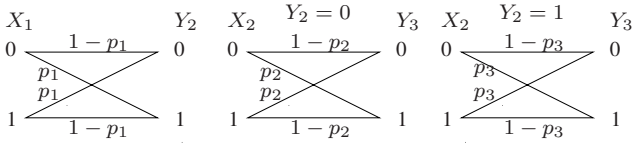}
\caption{Channel statistics of the degraded noncausal relay channel}
\label{ex:BSC}
\vspace{-.05in}
\end{figure}

When $p_1 = 0.2$, $p_2 = 0.1$, and $p_3 = 0.55$, we have
\begin{align*}
R_{\textrm{DF}} &= 0.2203, \\
R_{\textrm{CS}} &\ge 0.2566,\\
R_{\textrm{GP-DF}} &= C_{\infty} = R_{\textrm{NUB}} = 0.2453.
\end{align*}

The DF lower bound~\eqref{DecLwBnd} contains no auxiliary random variable and
thus can be computed easily. For the cutset upper bound~\eqref{CutUpBnd}, we
evaluate a simpler lower bound obtained by setting $X_2 = U$
\begin{align*}
 R_{\textrm{CS}} &\ge \max_{p(x_1)p(x_2|x_1,y_2)} \min \{I(X_1;Y_2),\;
I(X_2,X_1;Y_3)\}\\
&= 0.2566.
\end{align*}
In the capacity expression
(\ref{eq:capacity}) in Theorem
\ref{DgdCapacity}, the maximum is attained by $U \sim \mathrm{Bern}(1/2)$
independent of $(X_1,Y_2)$ and $X_2 = U \oplus Y_2$, which yields the capacity
$C_{\infty} = 0.2453$.

We prove this via a symmetrization argument motivated by Nair \cite{Nair2010}.
Note that
\begin{align}
C_{\infty} &= \max_{\substack{p(x_1)p(u|x_1,y_2)\\x_2(u,y_2)}} \min
\{I(X_1;Y_2),\;\nonumber\\[-12pt]
 &\hspace{9.2em}I(X_1,U;Y_3)-I(U;Y_2|X_1)\} \nonumber\\
&= \max_{p(x_1)} \min \Bigl\{I(X_1;Y_2), \nonumber\\
&\hspace{2em} \max_{\substack{p(u|x_1,y_2)\\x_2(u,y_2)}}(I(X_1,U;Y_3)-I(U;Y_2|X_1))
\Bigr\}.
 \label{eq:max}
\end{align}
Consider the maximum in the second term for a fixed $p(x_1)$. Assume without
loss of generality that $\Uc = \{1,2,\ldots,|\Uc|\}$. For any conditional pmf
$p_{U|X_1,Y_2}(u|x_1,y_2)$ and function $x_2(u,y_2)$, define $\Ut$, $\Xt_2$,
and $\Yt_3$ as
\begin{align}
p_{\Ut}(u) &= p_{\Ut}(-u)\nonumber\\
&= \frac{1}{2} p_{U}(u), \quad u\in \Uc,\nonumber\\
p_{X_1,Y_2|\Ut}(x_1,y_2|u) &= p_{X_1,Y_2|\Ut}(x_1,y_2|-u)\nonumber\\
&= p_{X_1,Y_2|U}(x_1,y_2|u), \quad u\in \Uc,\nonumber\\
\xt_2(u,y_2) &= 1-\xt_2(-u,y_2)\nonumber\\
 &= x_2(u,y_2), \quad (u,x_2) \in \Uc \times \{0,1\},\label{eq:x2}\\
p_{\Yt_3|\Xt_2,Y_2}(y_3|\xt_2,y_2) &= p_{Y_3|X_2,Y_2}(y_3|\xt_2,y_2),\quad y_3
\in \{0,1\}. \nonumber
\end{align}
Then for any $u \in \Uc$,
\begin{align}
p_{\Ut|X_1,Y_2}(u|x_1,y_2) &= p_{\Ut|X_1,Y_2}(-u|x_1,y_2) \nonumber\\
&= \frac{1}{2} p_{U|X_1,Y_2}(u|x_1,y_2), \label{eq:ut}\\
p_{Y_2|X_1,\Ut}(y_2|x_1,u) &= p_{Y_2|X_1,\Ut}(y_2|x_1,-u)\nonumber\\
&= p_{Y_2|X_1,U}(y_2|x_1,u),\\
p_{Y_3|X_1,\Ut}(y_3|x_1,u) &= 1-p_{Y_3|X_1,\Ut}(y_3|x_1,-u)\nonumber\\
&= p_{Y_3|X_1,U}(y_3|x_1,u)\label{SymU}.
\end{align}
Thus, $H(Y_2|X_1,U=u) = H(Y_2|X_1,\Ut = u)$ $ = H(Y_2|X_1,\Ut = -u)$ for all $u
\in \Uc$, which implies that
$H(Y_2|X_1,U) = H(Y_2|X_1,\Ut)$. Similarly, we can show $H(Y_3|X_1,U) =
H(\Yt_3|X_1,\Ut)$.
It can be also easily shown that for any $y_2 \in \{0,1\}$,
$p_{\Xt_2|Y_2}(0|y_2) = 1/2$, which implies that $p_{\Yt_3}(0) = 1/2$ and
$H(\Yt_3) = 1$.
Hence,
\begin{align*}
&I(X_1,U;Y_3)-I(U;Y_2|X_1)\\
& = H(Y_3)-H(Y_3|X_1,U)-H(Y_2|X_1)+H(Y_2|X_1,U)\\
& \leq H(\Yt_3) - H(\Yt_3|X_1,\Ut) - H(Y_2|X_1) + H(Y_2|X_1,\Ut) \\
& = \sum_{u = 1}^{|\Uc|} p_{U}(u)\,\big(H(Y_2|X_1,\Ut,|\Ut|=u)\\
&\hspace{2em}-H(\Yt_3|X_1,\Ut,|\Ut|=u)\big) +H(\Yt_3)-H(Y_2|X_1)\\
& \leq \max_{u\in \Uc} \big(H(Y_2|X_1,\Ut,|\Ut|=u)
-H(\Yt_3|X_1,\Ut,|\Ut|=u)\big)\\
&\hspace{2em}+ 1 - H(p_1),
\end{align*}
where the last maximum is attained by $p_{\Ut}(u) = p_{\Ut}(-u) = 1/2$ for a
single $u$. Note that from our definition of $\Ut$, this automatically
guarantees the independence between $\Ut$ and $(X_1,Y_2)$.
Therefore, the maximum in the second term of (\ref{eq:max}) is attained by $\Ut
\sim \mathrm{Bern}(1/2)$ independent of $(X_1,Y_2)$. Subsequently, we relabel
$\Ut$ as $U$ with alphabet $\{0,1\}$, $\Yt_3$ as $Y_3$, and $\Xt_2$ as $X_2$.

Now we further optimize the second term in (\ref{eq:max}), which we have
simplified as
\begin{align}
I(X_1,U;Y_3) - I(U;Y_2|X_1) &\stackrel{(a)}{=} I(X_1,U;Y_3) \nonumber\\
&\stackrel{(b)}{=} 1 - H(Y_3|X_1,U),
\label{eq:umax}
\end{align}
where $(a)$ follows by the optimal choice of $U$ that is independent of
$(X_1,Y_2)$ and
$(b)$ follows since $H(Y_3) = 1$. We maximize (\ref{eq:umax}) over all functions
$\xt_2(\ut,y_2)$ satisfying $x_2(0,y_2) = 1-x_2(1,y_2)$ for all
$y_2 \in \{0,1\}$. By the symmetry of $U$ as described in (\ref{SymU}),
$H(Y_3|X_1,U=0) = H(Y_3|X_1,U=1)$. Thus,
\begin{align*}
&H(Y_3|X_1,U)\\
&= p_U(0)H(Y_3|X_1,U=0) + p_U(1)H(Y_3|X_1,U=1)\\
 & = H(Y_3|X_1,U=0)\\
 & = p_{X_1}(0)H(Y_3|X_1=0,U=0)\\
 &\hspace{1em}+p_{X_1}(1)H(Y_3|X_1=1,U=0).
\end{align*}
By considering all four functions $x_2(u=0,y_2) \in \big\{\{0,1\}
\rightarrow \{0,1\}\big\}$ and removing the redundant choices by the symmetry of
the binary entropy function, we have
\begin{align*}
&H(Y_3|X_1=0,U=0)\\
&\geq \min\{H(p_1\bar{p}_2+\bar{p}_1\bar{p}_3),H(p_1\bar{p}_2+\bar{p}_1p_3)\}.
\end{align*}
Similarly,
\begin{align*}
&H(Y_3|X_1=1,U=0)\\
&\geq \min \{H(\bar{p}_1\bar{p}_2+p_1\bar{p}_3),H(\bar{p}_1\bar{p}_2+p_1p_3)\}.
\end{align*}
When $p_1 = 0.2$, $p_2 = 0.1$, and $p_3 = 0.55$, the minimum is attained by $X_2
= U \oplus Y_2$ for both terms regardless of $p(x_1)$. Therefore the second term
in (\ref{eq:max}) simplifies to
\[
1-p_{X_1}(0)H(p_1\bar{p}_2+\bar{p}_1p_3)-p_{X_1}(1)H(\bar{p}_1\bar{p}_2+p_1p_3).
\]
Finally, maximizing
\begin{align*}
&\min\{I(X_1;Y_2),\\
&\hspace{1.5em}1-p_{X_1}(0)H(p_1\bar{p}_2+\bar{p}_1p_3)-p_{X_1}(1)H(\bar{p}
_1\bar{p}_2+p_1p_3)\}
\end{align*}
over $p(x_1)$, we obtain the capacity $C_{\infty} = 0.2453$.
\end{example}

\appendices
\section{GP-PDF-CF lower bound recovers GP-DF lower bound}
\label{AppendixA}
We first establish an equivalent GP-DF achievable rate $R_{\textrm{GP-DF}}'$ and
an equivalent GP-PDF-CF achievable rate $R_{\textrm{GP-PDF-CF}}'$. Then we show
the equivalent GP-PDF-CF lower bound recovers the equivalent GP-DF lower bound.
\begin{claim}
The GP-DF lower bound (\ref{DFGP}) is equivalent to the following lower bound:
\begin{align*}
C_{\infty} &\ge R_{\textrm{GP-DF}}'\\
&= \max \min \{I(X_1;Y_2);
I(X_1,U;Y_3)-I(U;Y_2|X_1)\},
\end{align*}
where the maximum is over all pmfs $p(x_1)p(u|x_1,y_2)$ and functions
$x_2(x_1,u,y_2)$ such that $I(U;Y_3|X_1) \geq I(U;Y_2|X_1)$.
\end{claim}
\begin{IEEEproof}
Note that this lower bound differs from the GP-DF lower bound (\ref{DFGP}) in
that there is an additional constraint $I(U;Y_3|X_1) \geq I(U;Y_2|X_1)$ on the
input pmfs. Clearly, $R_{\textrm{GP-DF}}' \leq R_{\textrm{GP-DF}}$. We need to
show $R_{\textrm{GP-DF}}' \geq R_{\textrm{GP-DF}}$.

Let $X_1^*, X_2^*, U^*, Y_2^*,$ and $Y_3^*$ denote the random variables
evaluated at the optimum pmf of $R_{\textrm{GP-DF}}$. Let
$p_{X_2^*|X_1^*,Y_2^*}(x_2|x_1,y_2)$ be the conditional pmf induced by the
optimum pmf of $R_{\textrm{GP-DF}}$. If the constraint $I(U;Y_3|X_1) \geq
I(U;Y_2|X_1)$ is satisfied at the optimum pmf of $R_{\textrm{GP-DF}}$, then the
two lower bounds coincide. Now suppose at the optimum pmf of
$R_{\textrm{GP-DF}}$, the opposite is true, i.e., $I(U^*;Y_3^*|X_1^*) <
I(U^*;Y_2^*|X_1^*)$. Choose $X_1 = X_1^*$, $U = (X_1^*,W)$, where $W$ is
independent of $(X_1,Y_2)$, and $p_{X_2|X_1,Y_2}(x_2|x_1,y_2) =
p_{X_2^*|X_1^*,Y_2^*}(x_2|x_1,y_2)$ as the input pmf of $R_{\textrm{GP-DF}}'$.
Here the choice of $X_2$ is valid by the functional representation
lemma~\cite[Appendix B]{Kim}. Note that the constraint $I(U;Y_3|X_1) \geq
I(U;Y_2|X_1)$ is satisfied since it reduces to $I(W;Y_3|X_1) \geq I(W;Y_2|X_1) =
0$ by the independence between $W$ and $(X_1,Y_2)$. Hence,
\begin{align*}
&R_{\textrm{GP-DF}} \\
&= \min\{I(X_1^*;Y_2^*),
I(X_1^*,U^*;Y_3^*)-I(U^*;Y_2^*|X_1^*)\}\\
& \stackrel{(a)}{\leq} \min \{I(X_1^*;Y_2^*), I(X_1^*;Y_3^*)\}\\
& \stackrel{(b)}= \min \{I(X_1;Y_2), I(X_1;Y_3)\}\\
& \stackrel{(c)}{\leq} \min \{I(X_1;Y_2), I(X_1,W;Y_3)-I(W;Y_2|X_1)\}\\
& = \min \{I(X_1;Y_2), I(X_1,U;Y_3)-I(U;Y_2|X_1)\}\\
& \leq R_{\textrm{GP-DF}}',
\end{align*}
where $(a)$ follows since $I(U^*;Y_3^*|X_1^*) < I(U^*;Y_2^*|X_1^*)$, $(b)$
follows since the joint pmf $p(x_1,y_2,x_2)$ is preserved by our choice of $X_1$
and $X_2$ and thus the joint pmf $p(x_1,y_3)$ is preserved, and $(c)$ follows
since $W$ and $(X_1,Y_2)$ are independent.
\end{IEEEproof}

\begin{claim} 
The GP-PDF-CF lower bound (\ref{PDFCFGP}) is equivalent to the following lower
bound:
\begin{align}
&C_{\infty} \geq R_{\textrm{GP-PDF-CF}}'\nonumber\\
&= \max \min \{I(V,U;Y_3)+I(X_1;U,Y_3|V)-I(U;Y_2|V),\nonumber\\
&\hspace{5.5em}I(V;Y_2)+I(X_1;U,Y_3|V), \nonumber\\
&\hspace{5.5em} I(V;Y_2)+I(X_1;U,Y_3|V)\nonumber\\
&\hspace{6em}+I(U;Y_3|V)-I(U;Y_2|V)\},
\label{PDFCFGP2}
\end{align}
where the maximum is over all pmfs $p(v,x_1) p(u|v,y_2)$ and functions $
x_2(u,v,y_2)$ such that $I(X_1;Y_3|V)+I(U;X_1,Y_3|V) \ge I(U;Y_2|V)$. 
\end{claim}
\begin{IEEEproof}
Note that this lower bound differs from the origianl bound (\ref{PDFCFGP}) in
that there is an additional constraint $I(X_1;Y_3|V)+I(U;X_1,Y_3|V) \ge
I(U;Y_2|V)$ on the input pmfs. Clearly $R_{\textrm{GP-PDF-CF}}' \le
R_{\textrm{GP-PDF-CF}}$. We want to show $R_{\textrm{GP-PDF-CF}}' \ge
R_{\textrm{GP-PDF-CF}}$. 

Similarly let $X_1^*, X_2^*, U^*, V^*, Y_2^*,$ and $Y_3^*$ denote the random
variables evaluated at the optimum pmf of $R_{\textrm{GP-PDF-CF}}$. Let
$p_{X_2^*|V^*,Y_2^*}(x_2|v,y_2)$ be the conditional pmf induced by the optimum
pmf of $R_{\textrm{GP-PDF-CF}}$. If at the optimum pmf of
$R_{\textrm{GP-PDF-CF}}$, the constraint is satisfied, i.e.,
$I(X_1^*;Y_3^*|V^*)+I(U^*;X_1^*,Y_3^*|V^*) \ge I(U^*;Y_2^*|V^*)$, then the two
bounds coincide. Now suppose at the optimum pmf of $R_{\textrm{GP-PDF-CF}}$, the
opposite is true, i.e., $I(X_1^*;Y_3^*|V^*)+I(U^*;X_1^*,Y_3^*|V^*) <
I(U^*;Y_2^*|V^*)$. Choose $V = V^*, X_1 = X_1^*, U = (V^*,W)$, where $W$ is
independent of $(V, Y_2)$, and $p_{X_2|V,Y_2}(x_2|v,y_2) =
p_{X_2^*|V^*,Y_2^*}(x_2|v,y_2)$ as the input pmf of $R_{\textrm{GP-PDF-CF}}'$.
Again the choice of $X_2$ is valid by the functional representation lemma. Note
that the constraint is satisfied for it reduces to $I(X_1;Y_3|V)+I(W;X_1,Y_3|V)
\ge I(W;Y_2|V) = 0$ by the independence between $W$ and $(V,Y_2)$. 
Hence,
\begin{align*}
R_{\textrm{GP-PDF-CF}} &\stackrel{(a)}{\leq}
\min\{I(V^*;Y_2^*),\,I(V^*;Y_3^*)\}\\
& \stackrel{(b)}{=} \min\{I(V;Y_2),\,I(V;Y_3)\}\\
& \leq \min\{I(V,W;Y_3)+I(X_1;W,Y_3|V),\\
&\hspace{3.4em}I(V;Y_2)+I(X_1;W,Y_3|V)\}\\
& \leq R_{\textrm{GP-PDF-CF}}'
\end{align*}
where $(a)$ follows since $I(X_1^*;Y_3^*|V^*)+I(U^*;X_1^*,Y_3^*|V^*) <
I(U^*;Y_2^*|V^*)$ and $(b)$ follows since the pmf $p(v,x_1)p(x_2|v,y_2)$ is
preserved by our choice of $(V,X_1,X_2)$ and thus the joint pmfs $p(v,y_2)$ and
$p(v,y_3)$ are preserved.
\end{IEEEproof}

Finally, taking $V = X_1$ reduces $R_{\textrm{GP-PDF-CF}}'$ to
$R_{\textrm{GP-DF}}'$, which completes the proof.

\section*{Acknowledgments}
\label{Acknowledgments}
The authors wish to express their gratitude to Young-Han Kim for stimulating
discussions and insightful guidance throughout the work. They are also grateful
to Paolo Minero for discussions on the hybrid coding techniques.


\bibliographystyle{IEEEtran}
\bibliography{InfoTheory}

\begin{IEEEbiographynophoto}{Lele Wang} 
(S'11--M'16) received the B.E. degree from Tsinghua University in 2009 and the Ph.D. degree from University of California, San Diego (UCSD) in 2015, both in Electrical Engineering. She is currently a joint postdoctoral fellow at Stanford University and Tel Aviv University. Her research focus is on information theory, coding theory, and communication theory. She is a recipient of the 2013 UCSD Shannon Memorial Fellowship and the 2013--2014 Qualcomm Innovation Fellowship.
\end{IEEEbiographynophoto}

\begin{IEEEbiographynophoto}{Mohammad Naghshvar} 
(S'07--M'13) received the B.S. degree in electrical engineering from Sharif University of Technology in 2007. He obtained the M.Sc. degree and the Ph.D. degree in electrical engineering (communication
theory and systems) both from University of California San Diego in 2009 and 2013, respectively. He is currently a senior R\&D engineer at Qualcomm Technologies Inc., San Diego, CA. His research interests include active learning and hypothesis testing, stochastic control and optimization, wireless communication and information theory.
\end{IEEEbiographynophoto}

\end{document}